\documentstyle[twoside,fleqn,espcrc2_pre,espcrc2,epsf]{article}

\newcommand{\p}{\mbox{$\underline{p}$}}

\newcommand{\leptons}{\mbox{$l^{+} \nu_{l}$} }

\begin{document}

\title{$B$ meson form factors from 
HQET simulations\thanks{
Presented by C.\ McNeile.}
}

\author{C.\  Bernard,\address{Department of Physics, 
Washington University, St.\ Louis, MO 63130, USA}
        C.\ DeTar,\address{Physics Department, University of Utah, 
        Salt Lake City, UT   84112, USA \\
        and Zentrum f\"ur interdisziplin\"are
        Forschung, Universit\"at Bielefeld, Bielefeld, Germany}
        Steven Gottlieb,\address{Department of Physics,
Indiana University, Bloomington, IN 47405, USA}
        U.\ M.\ Heller,\address{SCRI, The Florida State University,
Tallahassee, FL 32306-4130, USA}
        J.\ Hetrick,${}^a$
        B.\ Jegerlehner,${}^c$
        C.\ McNeile,${}^b$
        K.\ Rummukainen,\address{Fakult\"at f\"ur Physik,
Universit\"at Bielefeld, Bielefeld, Germany}
        R.\ Sugar,\address{Department of Physics,
University of California, Santa Barbara, CA 93106, USA}
        D.\ Toussaint\address{Department of Physics,
University of Arizona, Tucson, AZ 85721, USA}
        and
        M.\ Wingate\address{Physics Department,
University of Colorado, Boulder, CO 80309, USA} }

\begin{abstract}
We use simulations of heavy quark effective field
theory to calculate the Isgur-Wise function,
and we demonstrate the feasibility of calculating the 
matrix element for the $B \rightarrow \pi + \leptons$ decay in the
lattice heavy quark effective theory (HQET).
\end{abstract}
 
\maketitle

\section{INTRODUCTION}
We describe the calculation of the hadronic matrix elements that are
required for the extraction of the $V_{cb}$ and $V_{ub}$ CKM matrix
elements from experimental data \cite{Onogi97a0}.  To reach the bottom
quark mass our strategy is to interpolate between results from
relativistic quarks with $m_q \le m_c$ and results from lattice HQET
~\cite{Mandula92a}.  Here we discuss only the HQET simulations, as our
clover form factor simulations have only just started.

All of our simulations use $n_f = 2$ dynamical staggered configurations
with a volume $16^3\times48$ and $\beta = 5.445$.  


\section{ISGUR-WISE FUNCTION}

The Isgur-Wise function is the QCD matrix element required in the
extraction of $V_{cb}$ from experimental data.  Experimental
measurements of the slope of the Isgur-Wise function vary from
$0.31$ to $1.17$, and the variations in theoretical predictions are
nearly as large~\cite{Neubert97a}.  Initial attempts to calculate the
Isgur-Wise function in lattice HQET had problems either with the
signal to noise ratio~\cite{Mandula94a0}
 or the renormalization factors~\cite{Hashimoto96}. The first
complete calculation has been done recently by the Kentucky
group~\cite{Draper96a0}.

We use the same method as the Kentucky group (see 
also~\cite{Mandula94a0,Hashimoto96}). We ran at all permutations of the 
following velocities:  $(0,0,0)$,  $(0.1,0,0)$,  $(0.25,0,0)$ and  $(0.5,0,0)$.
Our sample size is $80$ configurations, and our Wilson $\kappa$ values are $0.160$
and  $0.163$. A relative smearing function of 
$e^{-0.67 r }$ was used between the quarks in the $B$ meson.
\begin{figure}
\vbox{\epsfxsize=2.7in \epsfbox{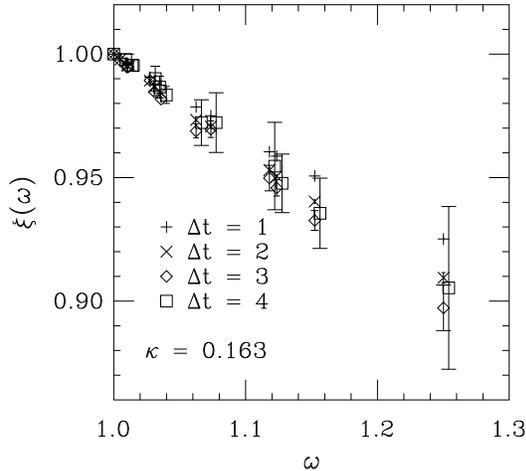}}
\vskip -9mm
\caption{Unrenormalized Isgur-Wise function}
\label{fig:isgurwisefunc}
\vskip -8mm
\end{figure}
In Fig.~\ref{fig:isgurwisefunc} we plot the bare Isgur-Wise function
for various time separations between the current and the $B$ source.
If the ground state has been isolated, then the Isgur-Wise function
should be independent of this separation. The data for $\Delta t =
2,3,4$ are consistent within present errors.

It is traditional to report the slope of the Isgur-Wise
function as a function of the 
dot product of the two meson velocities ($\omega =  v \cdot u $).
In Table~\ref{tb:rhoLINEAR} we report fits to 
\begin{equation}
\xi(\omega) = 1 - \rho^{2}( \omega -1 )
\label{eq:fitmodelSLOPE}
\end{equation}

\begin{table}
\begin{tabular}{|c|c|c||c|c|} \hline
$\Delta t$ & \multicolumn{2}{c||}{$\kappa = 0.160 $}  &
              \multicolumn{2}{c|}{$\kappa = 0.163 $}  \\ \cline{2-5}
 & $\rho^2$ & $\chi^2/df$  &$\rho^2$ &  $\chi^2/df $ \\ \hline
  $2$      &    $0.415(8)$  & $200/44 $   & $0.412(9)$ & $171/44$  \\
  $3$      &    $0.48(4)$  & $31/44$   &    $0.47(4)$  & $24/44$ \\
  $4$      &    $0.48(12)$ & $18/44$   &    $0.48(13)$ & $17/44$  \\ \hline
\end{tabular}
\caption{Preliminary fits to bare Isgur-Wise function data}
\label{tb:rhoLINEAR}
\end{table}
Assuming negligible quark mass dependence, 
our best estimate is therefore $\rho^2 = 0.48(13)$ at the physical light quark
mass. 
(We stress that it can not be compared with the experimental
value until the renormalization factors calculated in~\cite{Draper96a0} are included.)
For comparison, at $\beta = 6.0$ the
Kentucky group gets a bare $\rho^2 = 0.56$ and Hashimto and
Matsufuru~\cite{Hashimoto96} get $\rho^2 \sim 0.54$ (where we have
approximately removed the effect of tadpole improvement from their
result~\cite{Draper96a0}).

All the simulations of lattice HQET~\cite{Hashimoto96,Draper96a0} show a very
weak dependence of the slope on the light quark mass. However, the UKQCD
collaboration~\cite{Bowler95a} found a statistically significant
decrease in $\rho^2$ with light quark mass in their simulations that
used clover quarks for the $b$ quark. We also tried fitting our bare
Isgur-Wise function data to a
fit model that had quadratic corrections of $\omega$ in Eq.~\ref{eq:fitmodelSLOPE}.
Acceptable fits were found with approximately the same slope as in
Table~\ref{tb:rhoLINEAR} and positive curvature.




The velocity of the lattice HQET action is
renormalized~\cite{Aglietti94a}  because the action breaks Lorentz
symmetry. As we described last year~\cite{Bernard97b0}, we have tried
to estimate the renormalized velocity from the dispersion relation
of an HQET meson at finite residual momentum~\cite{Hashimoto96}.  The
renormalized velocity can be implicitly defined from
\begin{equation}
 E(\p,v^{R}) -E(0,v^{R}) =  \frac{\underline{v}^{R} .\p } {v_{0}^{R}}
\label{eq:HQETdisp}
\end{equation}
where $E(\p,v^{R})$ is the energy of the HQET meson at finite 
residual momentum ($\p$). In Table~\ref{tb:velRENORM}
we show the results of the non-perturbative 
velocity renormalization. We fit all the correlators in the 
time region 5 to 13 and obtained correlated $\chi^2/df $ slightly less than one.
For comparison, we also show the results for the perturbative
renormalization calculated by Mandula and Ogilvie~\cite{Mandula96,Mandula97}, using both a
boosted ($g^2 / u_0^4$) $PT_{boost}$ and bare coupling $PT_{bare}$, as well as using the
tree-level tadpole improved estimate~\cite{Draper96a0}.

The results in Table~\ref{tb:velRENORM} show that the velocity
renormalization is large.  These results would suggest that
perturbation theory with a boosted coupling agrees
best with the non-perturbative result. However, other analyses found
better agreement between the tree-level tadpole scheme~\cite{Draper96a0} and 
the non-perturbative calculations~\cite{Mandula96,Mandula97,Hashimoto96}.
This issue is under investigation.

\begin{table}
\begin{tabular}{|c|c|c|c|c|} \hline
Vel & NP & $PT_{bare}$ & $PT_{boost}$ & tadpole \\ \hline
$0.1$  & $0.04(2)$  & $0.074$  & $0.051$  & $0.085$  \\
$0.25$  & $0.13(2)$  & $0.18$  & $0.12$  & $0.21$  \\
$0.5$  & $0.25(3)$  & $0.34$  & $0.23$  & $0.41$   \\ \hline
\end{tabular}
\vspace{2pt}
\caption{Various estimates of the HQET velocity renormalization}
\label{tb:velRENORM}
\end{table}

\section{$B \rightarrow \pi $ FORM FACTOR}

The observation of the decay $B \rightarrow \pi + \leptons $ allows a
determination of $V_{ub}$, if the relevant QCD form factors can be
calculated.  There have been a number of lattice QCD calculations of
the required form factors (see~\cite{Onogi97a0,Flynn97a0} for
reviews). However, previous approaches suffer from the drawback that
calculations are done at large $q^2$, thus requiring a large
extrapolation to $q^2 \approx 0$, where measurements are currently
made.  To reach a low $q^2$ requires large meson velocities---not
easily achieved for heavy mesons in the NRQCD- or
propagating-quark-approaches.  As we have shown~\cite{Bernard97b0}, a
good signal can be obtained for the HQET $B$ meson with a large
velocity ($ v \approx 0.8$), so we propose the use of HQET---light
simulations to explore lower $q^2$ (see~\cite{Sloan97a0} for similar
ideas). It is not clear that HQET will be a good approximation to the
dynamics of the $B$ meson, at these values of $q^2$, nor that a
sufficiently good signal will be obtained.  However, results should
be very useful in helping to reduce the heavy quark extrapolation
errors over simulations that only use clover quarks.

\begin{figure}
\vbox{\epsfxsize=2.7in \epsfbox{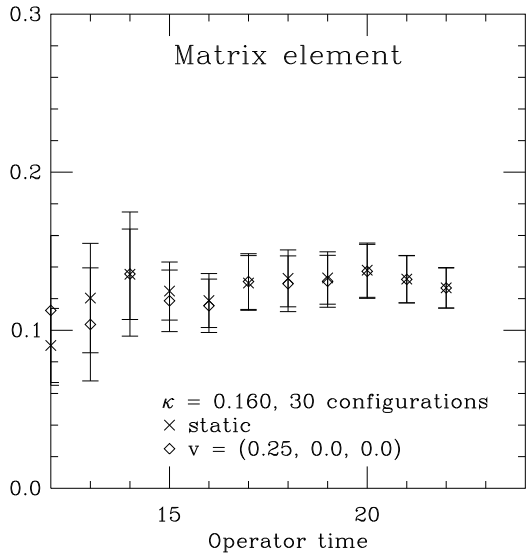}}
\vskip -0.3in
\caption{HQET---light matrix element}
\label{fig:hqetLIGHT}
\vskip -0.3in
\end{figure}

Because $B \rightarrow$ light meson form factors have never been studied before
using lattice HQET (although the static limit was studied in~\cite{UKQCD95c}),
we have computed the matrix element for 
$B \rightarrow \pi + \leptons$ using
HQET to check for a signal.
We use the setup described in~\cite{UKQCD95c} with the heavy clover
quark replaced by a HQET quark.
In Fig.~\ref{fig:hqetLIGHT} we plot the 
ratio of three point functions to two point functions
\begin{equation}
\frac{ \langle C_{3}(t;t_f)  \rangle } 
      { \langle C_{2}(t)_{L} \rangle   \langle C_{2}(t_f - t)_{B} \rangle } 
\label{eq:matelemeDEFN}
\end{equation} 
 that is
proportional to the $ \langle B \mid J_{\mu} \mid \pi \rangle$
matrix element, as a
function of the operator time $t$. The $B$ source is fixed at $t_f =
23$,
and the light meson source is fixed at $t=0$.

%
%

This work is supported by the DOE and the NSF.  The computations were
carried out at CCS (ORNL).  This presentation was prepared with the
kind additional support and assistance of the University of Bielefeld
and its Zentrum f\"ur Interdisziplin\"are Forschung.


\end{document}